\documentclass[aps,prl,twocolumn,showpacs,superscriptaddress,reprint]{revtex4-1}

\pdfoutput=1
\usepackage{amsmath,amssymb,xspace,graphicx,xcolor}

\usepackage[pdftex]{hyperref}
\newcommand{\eps}{\varepsilon}

\begin{document}

\title{From Phase to Micro-Phase Separation in Flocking Models:\\ The
Essential  Role of Non-Equilibrium Fluctuations}

\author{Alexandre P. Solon}
\affiliation{Universit\'e Paris Diderot, Sorbonne Paris Cit\'e, MSC, UMR 7057 CNRS, 75205 Paris, France}
\author{Hugues Chat\' e}
\affiliation{Service de Physique de l'\'Etat Condens\'e, CNRS URA 2464, CEA-Saclay, 91191 Gif-sur-Yvette, France}
\affiliation{LPTMC, CNRS UMR 7600, Universit\'e Pierre \& Marie Curie, 75252 Paris, France}
\affiliation{Beijing Computational Science Research Center, 3 Heqing Road, Haidian District, Beijing 100080, China}
\author{Julien Tailleur}
\affiliation{Universit\'e Paris Diderot, Sorbonne Paris Cit\'e, MSC, UMR 7057 CNRS, 75205 Paris, France}

\date{\today}
 
\begin{abstract}
  We show that the flocking transition in the Vicsek model is best
  understood as a liquid-gas transition, rather than an order-disorder
  one. The full phase-separation observed in flocking models with
  $Z_2$ rotational symmetry is however replaced by a micro-phase
  separation leading to a smectic arrangement of traveling ordered
  bands. Remarkably, continuous deterministic descriptions do not
  account for this difference, which is only recovered at the
  fluctuating hydrodynamics level. Scalar and vectorial order
  parameter indeed produce different types of number fluctuations,
  which we show to be essential in selecting the inhomogeneous
  patterns. This highlights an unexpected role of fluctuations in the
  selection of flock shapes.
\end{abstract}


\maketitle

Many of the phenomena heretofore only invoked to illustrate the many facets
of active matter are now being investigated in careful experiments,
and more and more sophisticated models are built to account for them.
For flocking alone, by which we designate the collective motion of
active agents, spectacular results have been obtained on both
biological
systems~\cite{Bausch,Sumino,Dogic,Zhou,Cavagna,fish,fishHugues,Sokolov2007,Peruani2012}
and man-made self-propelled particles~\cite{Deseigne,rollers,
  Thutupalli}.  Nevertheless, it is fair to say that the current
excitation about flocking takes place while our understanding of the
simplest situations remains unsatisfactory.  This is true even for
idealized self-propelled particles interacting only via local
alignment rules, as epitomized by the Vicsek model
(VM)~\cite{Vicsek1995} which stands out for its minimality and
popularity. Twenty years after the introduction of this seminal model
for the flocking transition, and despite the subsequent extensive
literature~\cite{VicRev}, we still lack a global understanding of the
transition to collective motion.

\begin{figure}[tbp]
  \centering
    \includegraphics[scale=1.]{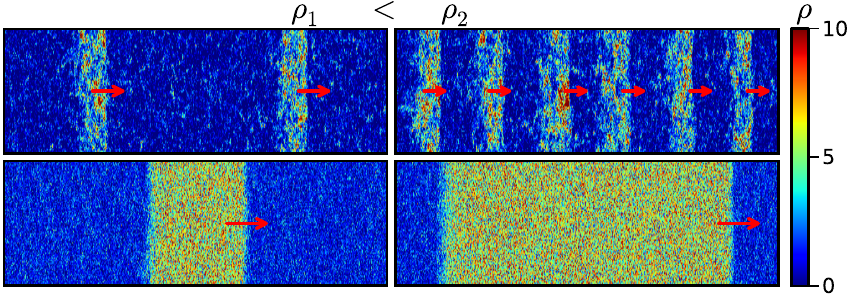}
    \caption{Top: Micro-phase separation in
      the Vicsek model. $\eta=0.4$, $v_0=0.5$, $\rho_1=1.05$ (left),
      $\rho_2=1.93$ (right). Bottom:  Phase separation in the Active Ising
      model. $D=1$, $\eps=0.9$, $\beta=1.9$, $\rho_1=2.35$ (left),
      $\rho_2=4.7$ (right).  System sizes $800\times 100$. Red arrows indicate
      the direction of motion.}
  \label{fig:domain_bands}
\end{figure}

It took a decade to show that the transition to collective motion in
the VM, initially thought to be critical~\cite{Vicsek1995}, was
discontinuous~\cite{GandH}: upon increasing the density or reducing
the noise strength, high-density bands of spontaneously aligned
particles form suddenly~\cite{GandH} (Fig~\ref{fig:domain_bands}).
The homogeneous ordered ``Toner-Tu'' phase~\cite{TT} is only observed
after a second transition at significantly lower noise/higher
density~\cite{GandH}. Since then, hydrodynamic-level deterministic
descriptions have been established and shown to support band-like
solutions \cite{BDG,Mishra2010,Ihle}, but it was recently proved
\cite{Caussin} that many such different solutions generically coexist.
In fact, the connection of these results to microscopic models remains
elusive.  More generally, we lack a unifying framework encompassing
the two transitions (disorder-bands, bands-Toner-Tu phase).

Such a global picture was recently proposed for the active Ising model
(AIM), where rotational invariance is replaced by a discrete symmetry
\cite{AIM}: particles carrying Ising spins diffuse in space
with a constant-amplitude bias along one arbitrarily fixed
direction $\pm {\bf u_x}$, the sign being given by the local
magnetization (see~\cite{Supp} for a detailed definition).  The
emergence of flocking in this model is akin to
a liquid-gas transition between an ordered liquid and a disordered
gas.  Unlike the traveling bands of the VM, inhomogeneous profiles in
the AIM are fully phase-separated, with a single macroscopic
liquid domain traveling in the gas
(Fig~\ref{fig:domain_bands}).
More generally, the symmetry difference between the two
models questions the relevance of this
framework for the VM.

In this Letter, we show that the flocking transition in the Vicsek
model is also best understood in terms of a liquid-gas
transition---rather than an order-disorder one---but with {\it micro
  phase} separation in the coexistence region. Closing a long-standing
debate, we indeed show that the dense ordered bands of the VM are
arranged periodically in space, leading to an effectively ``smectic''
phase.  We define an appropriate ``liquid fraction'' which allows us
to recover the usual linear behavior across the coexistence
region. But our most important results concern the hydrodynamic
descriptions of flocking models.  Surprisingly, deterministic
hydrodynamic equations for scalar (AIM) and vectorial (VM) order
parameter {\it both} support many different coexisting {\it stable}
inhomogeneous solutions selected by initial conditions, including
single-domain and micro-phase smectics. They thus do \text{not}
account for the differences observed in the microscopic models.  We
then show that the two scenarios are however discriminated at the
\textit{fluctuating} hydrodynamic level: the different symmetries
result in qualitatively different density fluctutations, effectively
providing a selection criterion. Scalar and vectorial stochastic
partial differential equations indeed generically lead to different,
unique profiles, in agreement with the microscopic models.

\begin{figure}[tbp]
  \centering
    \includegraphics[scale=1.]{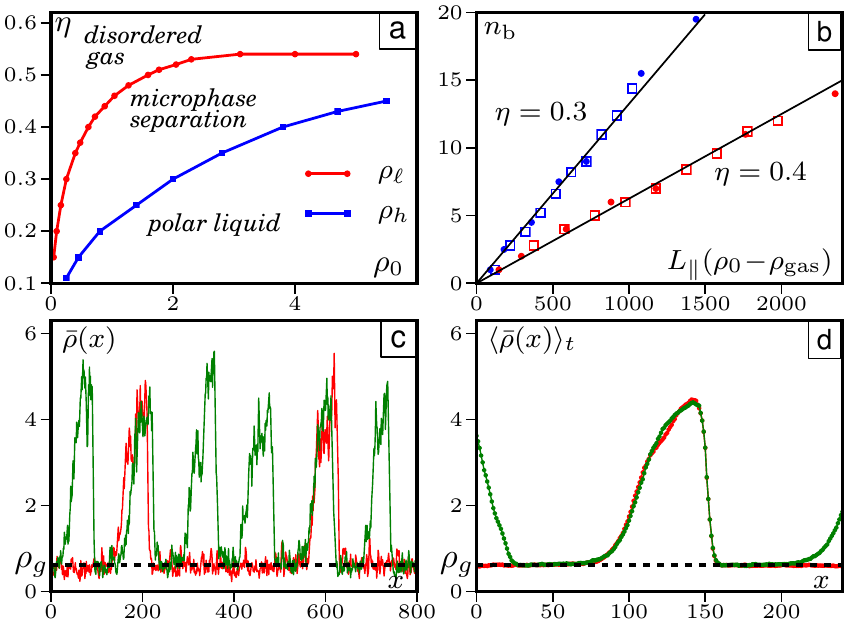}
    \caption{{\bf a}: Phase diagram of the Vicsek model. The binodals
      $\rho_{ \ell}(\eta)$ and $\rho_{ h}(\eta)$ mark the limit of the
      coexistence region. {\bf b}:~Number of bands vs $L_\parallel
      (\rho_0-\rho_{\rm gas})$ varying either the excess density for
      $Lx\times Ly=2000\times100$ (squares) or the system size along the
      direction of propagation (dots) for $\rho_0=0.6$ ($\eta=0.3$)
      or $\rho_0=1.2$ ($\eta=0.4$). The straight black lines are guide
      for the eyes. {\bf c}: Density profiles of
      Fig.~\ref{fig:domain_bands} averaged along the transverse
      direction ${\bf e}_\perp$, $\bar \rho(x_\parallel)=\langle
      \rho(x_\perp,x_\parallel)\rangle_{x_\perp}$. {\bf d}:
      Time average of the band profiles shown in {\bf
        c}. A three-fold increase of the excess density changes the
      number of bands but not the gas density or the shape of the
      bands.  $v=0.5$, $\eta=0.4$, $\rho_0=1.05$ (red), $\rho_0=1.93$
      (green).}
  \label{fig:phase_diagram}
\end{figure}

We first recall the definition of the Vicsek model. $N$ point-like
particles, labeled by index $i$, move at constant speed $v_0$ on a
rectangular plane of surface $S=L_xL_y$ with periodic boundary
conditions. At each discrete time step $\Delta t=1$, the headings
$\theta_i$ of all particles are updated in parallel according
to~\cite{foot_angular}
\begin{equation}
  \label{eq:rule_vicsek}
  \theta_i(t+1)=\langle \theta_j(t) \rangle_{j\in \mathcal{N}_i} +\eta\, \xi_i^t
\end{equation}
where $\mathcal{N}_i$ is the disk of unit radius around particle $i$,
$\xi_i^t$ a random angle drawn uniformly in
$[-\pi,\pi]$, and $\eta$ sets the noise intensity. Then,
particles hop along their new headings: ${\bf r}_i(t+1)= {\bf
  r}_i(t)+v_0 {\bf e}_i^{t+1}$, where ${\bf e}_i^{t+1}$ is the unit
vector pointing in direction given by $\theta_i({t+1})$.

In agreement with~\cite{GandH}, we find, varying the noise $\eta$ and
the density $\rho_0=N/S$, three different phases: a disordered gas at
high noise/low density, a polar liquid at low noise/high density, and
an intermediate region where ordered bands travel in a disordered
background. In the thermodynamic limit, the homogeneous phases are
separated from the coexistence phase by two ``binodals''
$\rho_{l}(\eta)$, $\rho_{h}(\eta)$, which are reported on
Fig~\ref{fig:phase_diagram}-a. One could in principle add spinodal
lines in the coexistence region, marking the limits of linear
stability of the homogeneous phases. At finite ``temperature'' $\eta$,
nucleation prevents us from computing them directly but quenching the
system into the coexistence region, we see two distinct behaviors:
metastability and nucleation close to the coexistence lines, spinodal
decomposition deeper in the coexistence region (see movies
in~\cite{Supp}). As for the AIM, an important difference with the phase
diagram of a liquid-gas phase transition in the canonical ensemble is
its unusual shape, which stems from the different symmetries of the
two phases. Since it is impossible to go continuously from the polar
liquid to the disordered gas, there is no supercritical region and the
critical point is sent to $\rho_c=\infty$.

While the phase diagrams of VM and AIM have similar shapes, their
coexistence regions are fundamentally different. Starting from random
initial conditions, the dynamics of the VM rapidly leads to {randomly}
spaced ordered bands propagating along a direction ${\bf e_\parallel}$
and spanning the system along ${\bf e_\perp}$, as reported
before~\cite{GandH,BDG}. On much longer time-scales, unreached in
these studies, the relaxation of compression modes actually leads to
\textit{regularly} spaced bands (See Fig.~\ref{fig:phase_diagram}-c
and movie in~\cite{Supp}).  In the thermodynamic limit, the bands have
well-defined profiles, {\it independent from the average density and
  the system size}. In this limit, increasing $\rho_0$ at constant
$\eta$ thus does not change the density $\rho_{\rm gas}$ of the
gaseous background, nor the celerity or the shape of the bands. Only
the band number $n_b$ increases, proportionally to $L_\parallel
(\rho_0-\rho_{\rm gas})$ (Fig.~\ref{fig:phase_diagram}-b\&d). For
finite systems, the quantization of the liquid fraction has some
interesting consequences. An excess mass $S (\rho_0-\rho_{\rm gas})$
which is not a multiple of the excess mass $m_b$ of a single band does
not allow the system to relax to its asymptotic band-shape. To
accomodate this excess mass the bands are slightly deformed, but $c$
and $\rho_{\rm gas}$ barely change as $\rho_0$ is varied (not shown).

\begin{figure}[tbp]
  \centering
    \includegraphics[scale=1.]{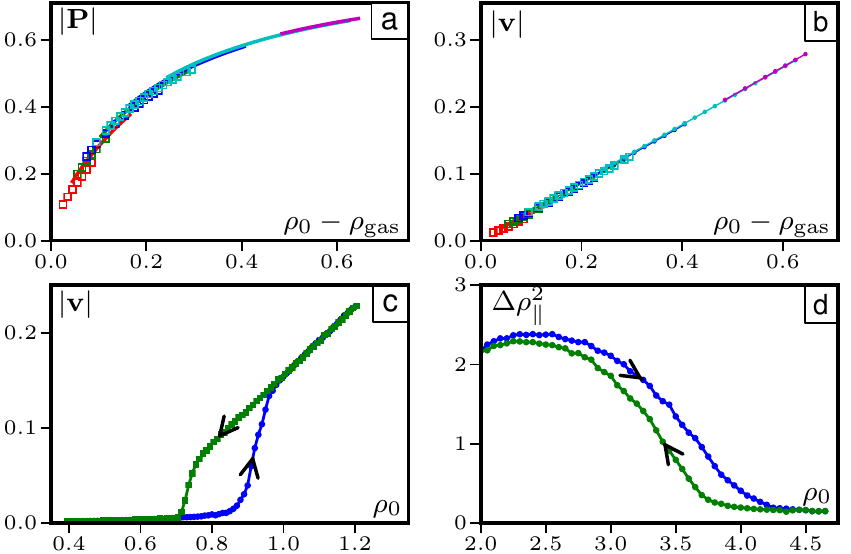}
    \caption{Polarisation ({\bf a}) and mean velocity ({\bf b}) vs
      $\rho_0$ for L=2048 (squares) and L=1024 (lines). Red, green,
      blue, cyan and magenta correspond to 1-5 band solutions. {\bf
        c}: Hysteresis loop between gas and micro-phase states. {\bf
        d}: Hysteresis loop between micro-phase and liquid states. The
      variance $\Delta\rho_\parallel^2$ quantifies inhomogeneity along
      the direction of motion. 100 runs are used for \textbf c and \textbf d,
with      $\eta=0.4$, system size $400\times 400$.}
  \label{fig:hysteresis}
\end{figure}

This smectic arrangement of finite-width bands markedly differs from
the more conventional liquid-gas phase-separation seen in the AIM,
where increasing the density simply widens the single liquid
domain. One may thus wonder whether all features of the liquid-gas
scenario survive. The global polarisation $|{\bf P}|=\frac 1 N |\sum_i
{\bf e_i}|$, used in previous studies of the VM to characterize the
onset of ordering~\cite{Vicsek1995,GandH,BDG}, does not show a linear
increase of the liquid fraction with density
(Fig~\ref{fig:hysteresis}-a). Such a scaling is however recovered by
considering $|{\bf v}|\equiv\frac 1 {S}|\sum_i v_0{\bf e_i}|=v_0\rho_0
|{\bf P}|$ (Fig~\ref{fig:hysteresis}-b). Indeed, for a propagating
band of celerity $c$, integrating the continuity equation $\dot \rho
=-\nabla \cdot {\bf W}$, where ${\bf W}({\bf r})=\sum_i \delta({\bf
  r}-{\bf r_i}) v_0{\bf e_i}$ is the momentum field, yields
$c(\rho({\bf r})-\rho_{\rm gas})= W_\parallel(\bf
r)$~\cite{BDG}. Averaging over space, this gives $| {\bf v}|=c
(\rho_0-\rho_{\rm gas})$. Since $c$ and $\rho_{\rm gas}$ barely depend
on $\rho_0$, $|{\bf v}|$ scales linearly with $\rho_0-\rho_{\rm gas}$,
even for finite systems where nearby values of $n_{\rm b}$ coexist
(See Fig~\ref{fig:hysteresis}-b). This is yet another signature of the
first-order nature of the transition and confirms the analogy with the
canonical liquid-gas transition; the apparent singularity of $|\textbf
P|$ close to $\rho_{\rm gas}$ is a simple consequence of its
normalisation, not of criticality (as often assumed in the
literature).

As expected~\cite{GandH}, we observe hysteresis loops when ramping
$\rho_0$ up and down close to $\rho_{ \ell}(\eta)$ with two sharp
jumps in the mean velocity $| {\bf v} |$
(Fig.~\ref{fig:hysteresis}-c). If the ramping is slow enough, they
correspond to the nucleation and vanishing of a single band which acts
as a critical nucleus. Indeed, a band can only be observed if the
excess density $\rho_0-\rho_{\rm gas}$ is of the order of $m_b/S$.  As
the system size increases, bands are hence seen closer and closer to
$\rho_{\rm gas}$ which thus coincide with the binodal $\rho_{\rm
  \ell}$, as in a standard liquid-gas transition. Moreover, the
critical nucleus contains a smaller and smaller fraction of the
particles as $L$ increases so that $|\textbf v|$ and $|\textbf P|$ vary {\it
  continuously} to zero in the infinite-size limit (cf.
Fig~\ref{fig:hysteresis}-a,b).

The second transition line $\rho_{\ell}(\eta)$ between the smectic
micro-phase and the ordered liquid is harder to locate accurately.
For $\rho_0\lesssim \rho_{ h}(\eta)$, the bands are indeed closely
packed and interact strongly. Although global orientational order
remains high, they break and merge in a chaotic manner (See 
movie in~\cite{Supp}). The resulting dynamics is thus difficult to
distinguish from the giant density fluctuations of the homogeneous
phase. Following~\cite{GandH}, we use $\Delta \rho_\parallel^2 \equiv
\langle (\bar \rho(x_\parallel) -\rho_0)^2 \rangle_{x_\parallel}$, the
variance along $L_\parallel$ of $\bar\rho$, the density profile
averaged in the transverse direction. Fig.~\ref{fig:hysteresis}-d
shows hysteresis loops of $\Delta \rho_\parallel$ around the
transition line. We then define $\rho_{\ell}(\eta)$ as the
high-density end-point of the loops.

\begin{figure}[tbp]
  \centering
    \includegraphics[scale=1.]{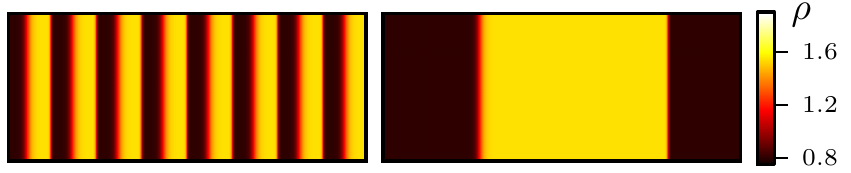}
    \caption{Density field in the PDEs after integration over
      $t=10^5$. {\bf Left}: scalar PDE, ordered initial condition with
      a periodic perturbation. {\bf Right}: vectorial PDE, disordered
      initial condition. Parameters:
      $\lambda=\rho_c=D=P_0=1$, $\rho_0=1.2$. System size
      $800\times100$.}
  \label{fig:sol_PDE}
\end{figure}

\begin{figure*}[tbp]
  \centering
    \includegraphics[scale=1.]{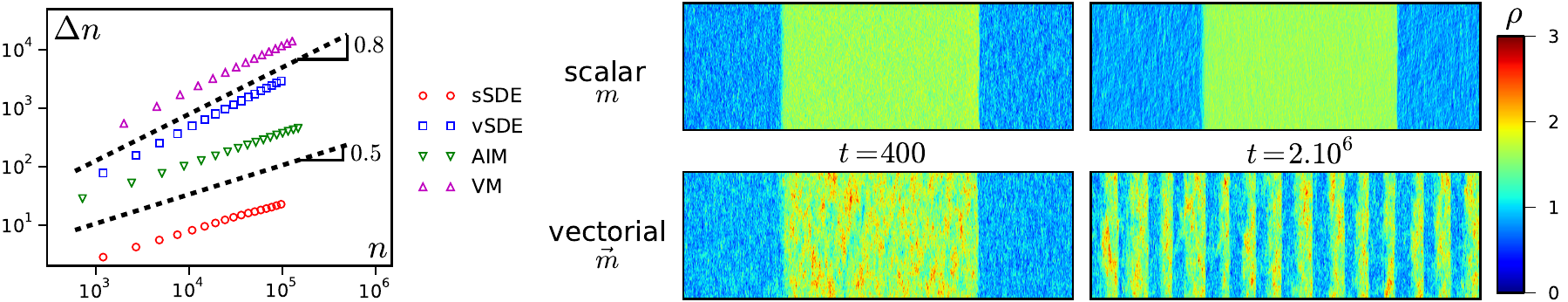}
    \caption{{\bf Left}: Number fluctuations $\Delta
      n=\sqrt{\langle n^2 \rangle -\langle n \rangle^2}$ where $n$ is
      the number of particles in boxes of different sizes. Measures
      done in the homogeneous liquid phase. Parameters: size $400\times 400$
      (all), $\rho_0=5$, $\beta=2.4$ (AIM), $\rho_0=5$, $\eta=0.4$
      (VM), $\lambda=\rho_c=D=P_0=1$, $\gamma^2=0.4$, $\rho_0=3$ (sSDE and
      vSDE). {\bf Right}:  Numerical integration of sSDE (top) and vSDE
      (bottom). Parameters: $\rho_c=\lambda=D=P_0=1$, $\gamma^2=0.4$,
      system size $2000\times 100$.}
  \label{fig:SDE}
\end{figure*}

To account for the differences between the coexistence phases of the
VM and AIM, we now connect the above results to the more theoretical
level of continuous descriptions. There are two important differences
between the hydrodynamic equations of VM and AIM: the nature of the
ordering field (vectorial in the VM, scalar in the AIM) and the
functional dependencies of the transport coefficients on density and
momentum fields. When looking for one-dimensional traveling solutions,
the dimension of the ordering field however becomes
irrelevant. Furthermore, it was recently shown~\cite{Caussin} that
hydrodynamic equations of flocking models admit such traveling
solutions with both smectic micro-phases and phase-separated
profiles. We have checked that both types of solutions exist for both
the equation proposed for the AIM~\cite{AIM} and for those proposed
for Vicsek-like models~\cite{BDG}.

Since~\cite{Caussin} only established the existence of these
solutions, a possibility to account for the different inhomogeneous
profiles seen in VM and AIM could be that these solutions have
different \textit{stability} in the corresponding two-dimensional
equations, where the dimension of the order parameter can play a
role. To test this hypothesis, we consider scalar and vectorial
versions of the ``same" minimal two-dimensional partial differential
equations (PDE). The first one, sPDE, has a scalar magnetization field
$W$ corresponding to the AIM discrete symmetry
\begin{eqnarray}
  \label{eq:sPDE1}
  \partial_t\rho &=& - \partial_x W \\
  \label{eq:sPDE2}
  \partial_tW &\!=\!& \!\Big[(\rho \!-\!\rho_{\rm t})\!-\! \frac{W^2}{P_0^2\rho}\Big] W  \!+\! \nu\nabla^2 W \!-\! \partial_x\rho
    \!-\! \lambda W\partial_x W
\end{eqnarray}
The second set, vPDE, has a vectorial momentum $\vec
W$ in line with the continuous rotational symmetry of the VM
\begin{eqnarray}
  \label{eq:vPDE1}
\!\!\!\!\!  \partial_t\rho &\!=\!& - \nabla\cdot \vec{W} \\
  \label{eq:vPDE2}
\!\!\!\!\!  \partial_t \vec{W} &\!=\!& \!\Big[\!(\rho \!-\!\rho_{\rm t})\!-\! \frac{|\vec{W}|^2}{P_0^2\rho}\Big] \vec{W}  \!+\! \nu\nabla^2 \vec{W} \!-\! \!\nabla\!\rho
    \!-\! \lambda (\vec{W}\!\cdot\!\nabla) \vec{W}
\end{eqnarray}
Clearly, the disordered solution $|W|=0$ becomes linearly
unstable for $\rho_0\!>\!\rho_{\rm t}$.  As in all active matter
systems with metric interactions, the homogeneous ordered solution
$|W|^2=\rho_0(\rho_0-\rho_{\rm t})P_0^2$ that emerges from this
mean-field transition is itself linearly unstable to long wavelengths
until $\rho_0>\rho_{\rm s}$ \cite{BDG,Mishra2010,Caussin}.
Note that $\rho_{\rm t}$ and $\rho_{\rm s}$ correspond to the spinodal
lines mentioned above.  Most of the inhomogeneous solutions classified
in \cite{Caussin} exist in a $\rho_0$ range wider than $[\rho_{\rm
  t},\rho_{\rm s}]$.  It is possible to estimate $\rho_{\rm min}$ and
$\rho_{\rm max}$, the extremal values of $\rho_0$ between which
solutions exist. For instance, setting all parameters including
$\rho_{\rm t}$ to unity as in Fig.~\ref{fig:sol_PDE}, one finds
$\rho_{\rm min}\simeq 0.808$, $\rho_{\rm s}\simeq 1.25$, and
$\rho_{\rm max}\simeq 1.74$.

We integrated numerically these two sets of equations for various
parameter values inside and outside the $[\rho_{\rm t},\rho_{\rm s}]$
interval~ \cite{foot_PDE}.  After transients, we end up with
effectively one-dimensional solutions taking constant values along
${\bf e}_\perp$.  In all cases we found both smectic micro-phases and
phase-separated profiles.  Which solution is observed depends only on
the initial condition and not on the symmetry of the ordering field.
Fig.~\ref{fig:sol_PDE} shows a periodic solution in the sPDE and
a single traveling domain in the vPDE obtained for the same parameter
values, a striking evidence that the (deterministic) hydrodynamic
equations alone {\it cannot} explain the selection of different
patterns observed in microscopic models. This result was found 
robust to modifications of Eqs.~\eqref{eq:sPDE1}-\eqref{eq:vPDE2}.

We call sSDE and vSDE the stochastic versions of
Eqs.~(\ref{eq:sPDE1}-\ref{eq:vPDE2}) obtained by adding a zero-mean
scalar (or vectorial) Gaussian white noise of variance $\gamma^2
\rho(1-\frac{|W|^2}{\rho^2})$ in the $W$ (or $\vec{W}$)
equation~\cite{footnoise}.  Integrating first sSDE and vSDE in the
homogeneous liquid phase, we recover the same density fluctuations as
in the corresponding microscopic models (Fig.~\ref{fig:SDE} left):
normal fluctuations in sSDE, giant ones in vSDE (with the same scaling
as in microscopic models).  More importantly, we recover the correct
type of inhomogeneous profiles in each case, irrespective of the
initial conditions.  For instance, starting from a large liquid domain
as initial condition in both sets of equations with the same
parameters, we find that sSDE keeps this configuration while it breaks
down in vSDE, eventually leading to a periodic array of bands
(Fig.~\ref{fig:SDE} right).  In the converse experiment, starting from
a configuration with many bands, we observe initially merging events
in both cases but this process stops in vSDE, leading to an asymptotic
periodic state with a finite number of bands, while coarsening
proceeds for sSDE.

We conclude that fluctuations play an essential role in selecting the
phase-separated patterns.  Note that similar experiments performed in
microscopic models yield similar results. For instance, in the VM at
relatively high noise large liquid domains are metastable for a long
enough time to be observed before fluctuations break them and lead the
system to the smectic micro-phase state (see movie in~\cite{Supp}).
Giant density fluctuations break large liquid domains and arrest
band-coarsening while normal fluctuations do not.  Two different
scenarios emerge: In the active Ising class, magnetization is a scalar
quantity, density fluctuations are normal and the system undergoes
bulk phase separation. In the active XY or Vicsek class, magnetization
is vectorial, density fluctuations in the liquid are anomalously large
and drive the system to the micro-phase separated state.

To summarize, we have shown that the flocking transition in the Vicsek
model amounts to a micro-phase liquid-gas transition in the canonical
ensemble exhibiting metastability, hysteresis and coexistence between
a disordered gas and a smectic arrangement of liquid bands.  This is
in contrast with the bulk phase separation exhibited by the active
Ising model~\cite{AIM}. We found that while (deterministic)
hydrodynamic equations do {\it not} explain this difference, their
stochastic counterparts do: the different nature of the order
parameter produces different types of number fluctuations, which are
essential in selecting the phase-separated patterns. This unexpected
role of fluctuations in the selection of flock shapes calls for a
greater care when trying to account for active systems based on purely
deterministic continuum equations.

Interesting questions remain open. For example, the mechanism by which
the bands interact in the VM to reach a periodic spacing and the
chaotic behavior of closely packed bands are still to be
investigated. Further, we so far have no analytical approach and
limited numerical results to ascertain the stability of the smectic
pattern in the direction along the bands.  It is not inconceivable
that, like recently found in active nematics~\cite{Sandrine}, the
coexistence phase is asymptotically disordered. Last, in the large
density region, the finite sizes of real flocking agents are not
negligible and steric effects such as motility-induced phase
separation~\cite{Tailleur2008,Fily2012,Redner2012} could enrich the
simple liquid-gas scenario~\cite{Farrell2012}.

\begin{acknowledgments}
  We thank the Max Planck Institute for the Physics of Complex
  Systems, Dresden, Germany and the Kavli Institute for Theoretical
  Physics, Santa Barbara, USA, for hospitality.  This research was
  supported in part by the National Science Foundation under Grant
  No. NSF PHY11-25915.
\end{acknowledgments}

\pagebreak
\appendix{}

\section{2d Active Ising model~\cite{AIM}}
$N$ particles carrying a spin $\pm 1$ move and interact on a 2d
lattice, an arbitrary number of particles being allowed on each
site. 

Spins flip according to an on-site ferromagnetic interaction: the spin
$S_k$ on site $i$ reverses its sign with rate
\begin{equation}
  \label{eq:ising_interaction}
  W(S_k\to -S_k)=\exp{\big( -\beta S_k\frac{m_i}{\rho_i} \big)}
\end{equation}
where $m_i$ and $\rho_i$ are the local magnetization and the number of
particles on site $i$, and $\beta$ plays the role of an inverse
temperature.

Particles can also hop to a neighboring lattice site: particle $k$, of spin $S_k$, hops
to the right (resp. left) with rates $D(1+\varepsilon S_k)$
(resp. $D(1-\varepsilon S_k)$) and to the top and bottom with rate
$D$. This sets an effective self-propulsion at speed $2D\varepsilon
S_k$ in the horizontal direction.

All simulations have been run using a random sequential update
Monte-Carlo scheme with a finite time-step
$dt=\frac{1}{N(4D+\exp(\beta))}$.

\section{Movies}
All movies are numerical simulations of the Vicsek
model~\cite{Vicsek1995}, as described in the main text.
\begin{itemize}
\item
  \href{http://www.msc.univ-paris-diderot.fr/~solon/SI_microphase/SI1_compressions-modes.avi}{Supplementary
    movie 1}: the relaxation of compression modes leads to a periodic
  arrangment of bands. $\eta=0.25$, $\rho_0=0.5$.
\item
  \href{http://www.msc.univ-paris-diderot.fr/~solon/SI_microphase/SI2_quench-from-gas_nucleation.avi}{Supplementary
    movie 2}: quench from the gas to the smectic micro-phase region
  (nucleation). $\eta=0.4$, $\rho_0=0.9$.
\item
  \href{http://www.msc.univ-paris-diderot.fr/~solon/SI_microphase/SI3_quench-from-gas_spinodal.avi}{Supplementary
    movie 3}: quench from the gas to the smectic micro-phase region (spinodal
  decomposition). $\eta=0.4$, $\rho_0=2$.
\item
  \href{http://www.msc.univ-paris-diderot.fr/~solon/SI_microphase/SI4_quench-from-liquid_nucleation.avi}{Supplementary
    movie 4}: quench from the liquid to the smectic micro-phase region
  (nucleation). $\eta=0.4$, $\rho_0=3.5$.
\item
  \href{http://www.msc.univ-paris-diderot.fr/~solon/SI_microphase/SI5_quench-from-liquid_spinodal.avi}{Supplementary
    movie 5}: quench from the liquid to the smectic micro-phase region
  (spinodal decomposition). $\eta=0.4$, $\rho_0=2.8$.
\item
  \href{http://www.msc.univ-paris-diderot.fr/~solon/SI_microphase/SI6_band-chaos.avi}{Supplementary
    movie 6}: chaotic bands close to the transition line
  $\rho_h(\eta)$. $\eta=0.2$, $\rho_0=0.5$.
\item
  \href{http://www.msc.univ-paris-diderot.fr/~solon/SI_microphase/SI7_domain-breaking.avi}{Supplementary
    movie 7}: a large liquid domain is broken by the giant density
  fluctuations, driving the system to the micro-phase separated
  state. $\eta=0.4$, $\rho_0=2.25$.
\end{itemize}


\begin{thebibliography}{99}
\bibitem{foot_angular} This defines the angular noise dynamics of the
  Vicsek model. Vectorial noise dynamics do not introduce any
  qualitative difference~\cite{GandH}.
\bibitem{foot_PDE} For the PDEs and SDEs we used a pseudo-spectral
  integration scheme (linear terms are computed in Fourier space,
  non-linear ones in real space) using semi-implicit
  time-stepping. SDEs were integrated following It\={o}'s interpretation.
\bibitem{footnoise} This structure of the noise variance is one of the
  simplest that accounts for the observed fact that the amplitude of
  fluctuations decreases when polarization $|W|/\rho$ increases while
  respecting the symmetry of the ordering field. However, our results
  are not sensitive to this particular form and hold for simpler
  choices such as $\gamma^2\rho$ or even additive noise.
\end{thebibliography}
\end{document}